\documentclass[%
aip,
{revtex4-1}  
jmp,%
amsmath,amssymb,
reprint,%
12pt, onecolumn,doublespacing]{revtex4-1}
\usepackage{graphicx}
\usepackage{dcolumn}
\usepackage{bm}

\begin{document}

\title{SPH simulations of turbulence in fixed and rotating boxes in two dimensions with no-slip boundaries\\}
\affiliation{
School of mathematical sciences, Monash University, Clayton VIC 3800, Australia} 
\author{A. Valizadeh}
\email{alireza.valizadeh@monash.edu.}
\author{J.J. Monaghan}
 \email{joe.monaghan@monash.edu.}
\date{\today}

\begin{abstract}
{In this paper we study decaying turbulence in fixed and rotating boxes in two dimensions using the particle method SPH. The boundaries are specified by boundary force particles, and the turbulence is initiated by a set of gaussian vortices.  In the case of fixed boxes we recover the results of Clercx and his colleagues obtained using both a high accuracy spectral method and experiments.  Our results for fixed boxes are also in close agreement with those of Monaghan  \citep{Monaghan2011} and Robinson and Monaghan  \citep{RobinsonMonaghan2011} obtained using SPH.  A feature of decaying turbulence in no-slip, square, fixed  boundaries is that the angular momentum of the fluid varies with time because of  the reaction on the fluid of the viscous stresses on the boundary.  We find that when the box is allowed to rotate freely, so that the total angular momentum of box and fluid is constant, the change in the angular momentum of the fluid is a factor $\sim 500$ smaller than is the case for the fixed box, and the final vorticity distribution is different.  We also simulate the behaviour of the turbulence when the box is forced to rotate with small and large Rossby number, and the turbulence is initiated by gaussian vortices as before.  If the rotation of the box is maintained after the turbulence is initiated we find that in the rotating frame the decay of kinetic energy, enstrophy and the vortex structure is insensitive to the angular velocity of the box. On the other hand, If the box is allowed to rotate freely after the turbulence is initiated, the evolved vortex structure is completely different.   }
\end{abstract}


\maketitle

%

\section{\label{sec:lntro}Introduction\protect\\ }

Smoothed particle hydrodynamics is now widely used in computational fluid dynamics especially in problems involving breaking waves, and free surfaces disrupted by the impact of rigid bodies.  In many of these systems the flows are initially laminar but develop turbulence as they evolve. A common example is the simulation of a laboratory dam break which becomes turbulent after the head of the flow hits an end wall where it forms a return wave that plunges into the incoming fluid \citep{Colagrossietal2003}. In order to have confidence in the predicted evolution of the flow it is necessary to determine how accurately SPH can simulate turbulence.  

Early studies of turbulence using SPH made use of known sub-grid models applied to two dimensional flows. For example Shao and Gotah \citep{ShaoGotah2004} and Shao et al. \citep{Shaoetal2006} used a 2D  turbulent viscous stress based on the Smagorinsky model. They applied their model to the turbulence generated by a succession of cnoidal waves breaking on a linear ramp  and used extensive phase averaging and other approximations (for example multiplying by a factor 4/3  to bring their results into better agreement with the experiments that are in 3D). They  obtained results in reasonable agreement with experiment although  SPH simulation without the turbulence model gave very similar results. Dalrymple and Rogers \citep{DalrympleRogers2006} also used a Smagorinsky model for 2D turbulence. Violeau and Issa \citep{VioleauIssa2007} studied 2D turbulence and compared the  $k-\epsilon$, the  EARSM (explicit algebraic Reynolds stress model), and an SPH version of Large Eddy Simulation. The comparison was made for a dam break, but  dam breaks have the disadvantage that while they are initially nearly 2D, they become 3D when the fluid strikes the wall of the experimental tank.  In particular, the front of the fluid formed by the returning, plunging wave is no longer 2D.  Apart from this problem there was not enough detailed information to determine if a particular method was superior. For example, the experiments did not provide them with the  decay of the kinetic energy, the decay of the enstrophy,  the vortex structure or the velocity correlation functions.

Over the last few years experiments and detailed spectral method simulations have been applied to study 2D turbulence in fluids with no-slip boundaries.  These include the discovery of spontaneous spin-up \citep{Clercxetal1998}, experiments on 2D turbulence in a stratified fluid \citep{Clercxetal1999}, and the effects of solid boundaries \citep{Heijstetal2006}. These results provide a convenient framework within which to study SPH turbulence and this has been exploited by \citep{RobinsonMonaghan2011}, and \citep{Monaghan2011}  to test SPH without turbulence models. In addition Monaghan \citep{Monaghan2011} discusses and applies an SPH turbulence model for these problems. 

A key feature of turbulence in a square no-slip boundary is that the angular momentum is not constant, and may change abruptly from the initial state then decay (\citep{Clercxetal1999,Clercxetal1998,Heijstetal2006,Maassen2002}). The SPH simulations of   \citep{RobinsonMonaghan2011} predict the same qualitative results though the details depend on the initial state.  The SPH and spectral methods give results for the decay of the kinetic energy, and the enstrophy which are in satisfactory agreement.  There are differences between the two SPH codes since that of Monaghan and Robinson models the boundary with layers of fixed fluid particles, and uses a cubic spline kernel, while that of Monaghan uses boundary force particles, and one of the Wendland kernels. Robinson (private communication) found that the results for the decay of the kinetic energy converged more rapidly when the cubic spline kernel was replaced by the Wendland kernel. However, as we shall show, most of the features of the turbulence in a fixed box, for example the kinetic energy decay, the enstrophy decay, and the structure of the evolved vorticity field, are very similar.

In the present paper the study of SPH simulation of turbulence will be extended using the SPH code of Monaghan (2011) but without the turbulence.  We first confirm that the SPH simulation for the case of decaying turbulence in a no-slip square box converges and the spin-up is consistent with that found using the spectral theory.  Second we study  decaying turbulence when the box containing the fluid is allowed rotate under the surface stresses produced by the fluid. In this case the total angular momentum of the system of box and fluid is conserved.  Third we simulate turbulence when the fluid and box are in rigid rotation when the turbulence is  initiated. The evolution of the turbulence was then studied both when the box was forced to rotate at its initial angular velocity, and when it was allowed to be driven by the fluid stress.  This problem is related to turbulence in the earth's atmosphere, but a more complete discussion along the lines of the $\beta$-plane study of \cite{Krameretal2006}, will not be attempted.


\section{The equations of motion in two dimensions}
We consider an incompressible fluid is moving in two dimensions within a square boundary with no-slip boundary conditions.  It is convenient, especially when we give the boundary a mass and allow it to rotate, to refer to the boundary as a box. The acceleration equation for the fluid is 
\begin{equation}
\frac{d{\bf v}}{dt} = \frac{\partial {\bf v}}{\partial t} + ({\bf v} \cdot \nabla){\bf v} = -\frac{\nabla P}{\rho} + \frac{\mu}{\rho} \nabla^2 {\bf v}.
\label{contNavStokes}
\end{equation}
From this equation it is straightforward to show that the rate of change of the kinetic energy $E_K$ of the fluid is given by
\begin{equation}
\frac{d }{dt} \int_{\mathcal A}  \frac12 \rho {\bf v}^2 dA = - \mu \int_{\mathcal A} \omega^2 dA,
\label{ekdt}
\end{equation}
where $\omega {\hat {\bf z}} $ is the vorticity, and ${\hat {\bf z}} $ is a unit vector perpendicular to the plane of the fluid. The integration over a volume is equivalent in the present case to an integration over the two dimensions of the fluid, and is denoted by ${\mathcal A}$. Because of the boundary condition, we have used the fact that ${\bf v}$ is zero on the boundary. The rate of change of the vorticity is given by
\begin{equation}
\frac{d\omega}{dt} = - \nu  \nabla^2 \omega.
\label{vort}
\end{equation}
The total enstrophy varies with time according to 
\begin{equation}
\frac{d }{dt} \int _{\mathcal A}   \frac12 \omega^2 dA = - \nu \int_{\mathcal A} ( \nabla \omega)^2 dA + \nu  \int_{\mathcal B}  \omega (\nabla \omega \cdot {\bf n} ) ds,
\label{enstr}
\end{equation}
where $\nu$ is the kinematic viscosity, ${\bf n}$ is an outward unit vector, and the second integration is around the boundary. These equations, together with one further equation giving the time variation of the gradients of the vorticity, were used by \cite{Batchelor1969} to develop his celebrated argument for the existence of a turbulent energy spectrum associated with a vorticity cascade to shorter length scales.

The rate of change of total angular momentum $L$ of the two dimensional fluid is given by 
\begin{equation}
\frac{d }{dt} \int_{\mathcal A}  \rho\left ( ( {\bf r} \times {\bf v}) \cdot {\hat {\bf z}}\right ) dA =  \int_{\mathcal B} P {\bf r} \cdot {\bf ds} + \mu \int_{\mathcal {A} } {\bf r} \times \nabla^2 {\bf v} dA
\label{angmom}
\end{equation}
The second integration can be written in terms of a surface stress, but it is convenient to work with the form given noting that for an incompressible fluid
\begin{equation}
\nabla^2 {\bf v} = -\nabla \times \omega {\bf \hat z},  
\end{equation}
and for a two dimensional fluid
\begin{equation}
{\bf r} \times (\nabla \times {\omega \bf \hat z }) = \nabla({\bf r} \cdot \omega \hat z)  - {\bf r} \cdot \nabla \omega.
\end{equation}

Making use of these various  relations we find 
\begin{equation}
\frac{dL}{dt} = \int_{\mathcal{B}} P {\bf r} \cdot {\bf ds} + \int_{\mathcal{B}} \omega {\bf r} \cdot {\bf n} ds - 2 \int_{\mathcal{A} }\omega dA.
\end{equation}
The last term can be written as the circulation around the boundary and this vanishes because of the no-slip condition.

In this paper we also consider turbulence in a two dimensional fluid contained within a no-slip, square boundary rigidly rotating with angular velocity $\Omega {\bf \hat z}$. In this case it is useful to consider the equations of motion in a frame rotating with angular velocity $\Omega$. For this purpose, we let ${\bf v}$ denote the velocity in this rotating frame, and $\omega {\bf \hat z}$ its vorticity.  The acceleration equation is then
\begin{equation}
\frac{d{\bf v}}{dt}  = - 2\Omega {\bf \hat z}\times {\bf v} +  \Omega^2 {\bf r}  -\frac{\nabla P}{\rho} + \frac{\mu}{\rho} \nabla^2 {\bf v}.
\label{contNavStokesRot}
\end{equation}
where $d/dt$ denotes the derivative following the motion in the rotating frame.  By taking the curl of this equation we get the same equations for the rate of change of vorticity as for the non rotating box. The equations for the rate of change of the  total energy and total enstrophy, calculated in the rotating frame,  are also the same as for the non rotating box.  We believe it is therefore reasonable to expect that Batchelor's argument would apply when the boundary rotates rigidly throughout the simulation.  We find that this conjecture is correct.

We also study the case where the boundary is given a mass and a moment of inertia (we then call it a box)  and allowed to rotate freely under the stresses from the fluid from the moment the turbulence is initiated.  In this case the total angular momentum of the fluid and the box is constant. In addition, the stress felt by the fluid is less because the box moves in response to the stress on it from the fluid.  The effect is similar to a person attempting to walk on a platform that is free to move.  One consequence of this is that the change in the angular momentum of the fluid is a factor 1/500 less than when the box is fixed.    Our simulations also show that the form of the vortices that evolve is very different from that in the case of a fixed box, or a  box which  rotates rigidly throughout the simulation. There is no simple expression for the rate of change of the fluid energy and vorticity in this case because the dynamics of the fluid must include the effects of the motion of the box.

\section{SPH model}
\label{Sec:SPHmodel}
We consider a weakly compressible fluid with pressure $P$ a function of density $\rho$. Surface tension is neglected. The reader is assumed to be familiar with standard SPH as described in the reviews by \cite{Monaghan1992,Monaghan2005}.  In the following, the labels $a$ and $j$ are used for SPH fluid and boundary particles respectively, and $\eta$ is used when a summation is over both fluid and boundary particles. The SPH form of continuity equation is

\begin{equation}
\frac{d\rho_a}{dt} = \rho_a \sum_{\eta} \frac{m_\eta}{\rho_\eta}\left({\bf v}_a - {\bf v}_\eta \right) \cdot\nabla_aW_{a\eta} ,
\label{continuity}
\end{equation}
where the mass, position, velocity, density and pressure of particle $a$ are $m_a$, $r_a$, ${\bf v}_a$, $\rho_a$, and $P_a$, respectively. The summation is over all particles. The function $W_{a\eta }= W(| {\bf r}_{a\eta}|, h)$  is the SPH kernel, $| {\bf r}_{a\eta} |$ is the distance between particle $a$ and particle $\eta$, and $h=(h_a+h_\eta)/2$ is the average smoothing length. In the calculations to be described the kernel is the  fourth-order Wendland function \citep{Wendland1995} for two dimensions. This function, when normalized so that $2 \pi\int W(r,h) rdr = 1$, is given by
\begin{equation}
W(z,\ell) = \frac{7}{ 64 \pi h^2} (2-z/h)^4 (1+2z/h),
\end{equation}
if $z\le 2h$, and zero otherwise.   Simulations of a wide variety of problems show that the choice $h_\eta=1.5\delta$, where $\delta$ is the initial particle spacing, gives good results. The interaction between any two fluid particles is  zero beyond $3 \delta$. The gradient taken with respect to the coordinates of particle $a$ is denoted by $\nabla_a$. The pressure of fluid particle $a$ is given by
\begin{equation}
 P_a= \frac{\rho_0 c_s^2}{7}\left(\left(\frac{\rho_a}{\rho_0}\right)^7 - 1\right),
\label{pressure}
\end{equation}
where $\rho_0$  is the reference density of the fluid. The speed of sound $c_s$ is 10 times the maximum speed of fluid $V_{max}$ which we estimate from the initial velocity field.  The boundary particles have zero pressure.

The acceleration equation for the  SPH particle $a$ is
\begin{equation}
 \frac{d{\bf v}_a }{ dt} =-\sum _{\eta}m_\eta \left(\frac{P_{a}}{\rho_a^2}+\frac{P_{\eta }}{\rho_{\eta}^2} - \Pi _{a\eta} \right )\nabla_{a}W_{a\eta}+\sum_{j} m_{j}{\bf f}_{aj}.
\label{momentum1}
\end{equation}
The first summation in (\ref{momentum1}) is over all  particles and the second is over the boundary particles. The viscosity is determined by $\Pi _{a\eta}$ for which we use the form
\begin{equation}
\Pi _{a\eta}=-\alpha \frac{{\bar c} }{\bar{\rho}_{a\eta} } \frac{ {\bf v}_{a\eta}\cdot {\bf r}_{a\eta} }{|{\bf r}_{a\eta}|},
 \label{viscosity}
\end{equation}
where $\alpha$ is a constant, ${\bf v}_{a\eta}= {\bf v}_a-{\bf v}_\eta$, ${\bf r}_{a\eta}={\bf r}_a- {\bf r}_\eta$, ${\bar \rho}_{a\eta} =(\rho_a + \rho_\eta )/2$ denotes the average density, and $\bar{c}= (c_a + c_\eta)/2$. The constant $\alpha$ can be written in terms of the kinematic viscosity by converting the summations to integrals.  We find for the Wendland kernel  \citep{KajtarMonaghan2010} that 

\begin{equation}
 \nu=\frac{1}{8}\alpha h\overline{c}.
 \label{kineticviscosity}
\end{equation}

The last term in Eq.~(\ref{momentum1}) is the boundary force on fluid particle $a$ due to the boundary particles.  This force, together with the viscous forces due to the boundary particles included in the first term, is equivalent to the Sirovich \citep{Sirovich1967} formulation of the boundary conditions in terms of boundary forces, and closely related to the Immersed Boundary Method of Peskin \citep{Peskin1977}. ${\bf f}_{aj}$ is given by \citep{MonaghanKajtar2009}
\begin{equation}
 {\bf f}_{aj}=\frac{\Gamma\ \Phi_{aj}}{(|{\bf r_{aj}}|-\Delta)} \frac{{\bf r}_{aj} }{ |{\bf r}_{aj} |},
 \label{boundforce}
\end{equation}
where $\Phi_{aj}=\frac{1}{32}(1+\frac{5}{2}q+2q^2)(2-q)^5$ for $q\le 2$ and is otherwise 0.  $\Gamma$ is a constant equal to $2V_{max}^2/(m_a+m_j)$ , $\Delta = \delta/3$ is  the boundary particle spacing , and $q={|\bf r_{aj}|}/{h_{aj}}$.\\ 

The position of any fluid particle $a$ is found by integrating
\begin{equation}
\frac{d{\bf r}_a}{dt} = {\bf v}_a.
 \label{replacement}
\end{equation}

For convenience in describing the time stepping algorithm we write the equations in the form
\begin{equation}
\frac{d{\bf v}_a}{dt}={\bf F}_{a}({\bf r},\rho,{\bf v}),
 \label{momentum2}
\end{equation}
and
\begin{equation}
 \frac{d\rho_{a}}{dt}= {\bf D}_{a}({\bf r},{\bf v}).
 \label{continuity2}
\end{equation}

These equations were integrated using a time stepping scheme that is second order and based on Verlet symplectic method. In the following $A^0$ denotes a quantity $A$ at the beginning of the current time step, $A^{1/2}$ at the midpoint of the step, and $A^1$ at the end of the step. The time stepping equations, where $\delta t$ is the time step, can then be written
\begin{eqnarray}
{\bf r}_a^{1/2} &= &{\bf r}_a^0+\frac{\delta t}{2}{\bf v}_a^0,\\
 \label{displacement tstep}
 {\bf v}_a^{1/2}& = &\kappa \ ({\bf v}_a^0+\frac{\delta t}{2}{\bf F}_a^{0}),\\
 \label{velocity tstep}
 \rho_a^{1/2}&= &\rho_a^{0}+\frac{\delta t}{2}{\bf D}_a^0.
 \label{density tstep}\end{eqnarray}
where $\kappa$ is a damping  factor used to bring the particles to equilibrium before initializing the turbulence. This is required because the boundary forces are initially unbalanced. The simple and efficient method which we use here is that $\kappa$ is 1.0 for all steps except that every 4th time step it is set to 0.9854. This value of $\kappa$ is optimal to reduce the kinetic energy to the less than 0.01 percent of the total kinetic energy of vortices after 1000 damping steps. For each simulation a minimum time is necessary for damping, so that the number of damping steps will increase as resolution increases ($\delta$ made smaller).

The time step is completed by calculating  ${\bf v}_a^1$, ${\bf r}_a^1$, and $\rho_a^1$ according to
\begin{eqnarray}
{\bf v}_a^1&= &\kappa \ ({\bf v}_a^0+{\delta t}{\bf F}_a^{1/2} ),\\
 \label{velocity tstep2}
{\bf r}_a^1&= &{\bf r}_a^{1/2}+\frac{\delta t}{2}{\bf v}_a^1,\\
 \label{displacement tstep2} 
 \rho_a^{1}&= &\rho_a^{1/2}+\frac{\delta t}{2}{\bf D}_a^1 ({\bf r}_a^1,\rho_a^{1/2},{\bf v}_a^1).
 \label{density tstep2}
\end{eqnarray}

 To improve the speed we replace ${\bf F}_a^0$ in \ref{velocity tstep} by ${\bf F}_a^{-1/2}$, i.e half a step back. \\
SI units are used throughout this paper.

\section{The Initial turbulent velocity field}
\label{Sec:velocity field}

For the numerical study of 2D decaying turbulence in a container the initial velocity  field has been set using either Chebyshev polynomials \citep{Clercxetal1998,Clercxetal1999,Clercxetal2001} or Gaussian vortices  \citep{ClercxHeijts2000,ClercxNielsen2000}. The results for both setups are similar. In this paper Gaussian vortices are used. For the general case N$\times$N equal-size Gaussian vortices are placed on a regular lattice, in a checker-board pattern of positive and negative vorticity. The initial distance between the centres of vortices, $\lambda$ is $S/(N+1)$, where $S=1$ is the width of box. The centres of the vortices were then given a random displacement  $0.06\lambda (2\tau-1)$, where $\tau$ is a quasi random number between 0 and 1. The rule for the random number is  $jran=mod(jran*k+l,m)$ and $\tau=jran/m$ \citep{FORTRAN77}. Here $k$, $l$, and $m$ are 106, 1283 and 6075, respectively. When we wish to compare the effect of different initial states we use $12$ different initial values of $jran$. The values were $11$, $13$, $17$, $23$, $37$, $49$, $137$, $191$, $1111$, $1117$, $1139$, and $3737$. 

The SPH particles were placed on a grid of squares in the domain $0\le x \le 1$ and $0\le y \le 1$.  Each SPH fluid particle`s velocity was calculated by summation over the vortices according to \citep{Monaghan2011}: 

\begin{equation}
 \bar{v}_a =\sum_{k=1}^{N\times N}\textbf{r}_{a k}\frac{\omega_k a_k}{2\pi |{\bf r}_{a k}|^2} \left(1- e^{-|{\bf r}_{a k}|^2/a_k^2 }\right),
 \label{Gaussian vortices}
\end{equation}
where, ${\bf r}_k$ is the center of the kth vortex, $\omega_k$ is the vortex strength and its absolute amount is equal to $1.0$ for all vortices, and $a_k$ is the  amplitude of vortex $k$ equal to $S/{4N}$. Finally, to ensure that the no-slip condition is closely satisfied in the initial state, a smoothing function similar to that used by other authors  \cite{Clercxetal1998} was applied . The final initial velocity field for fluid particle $a$ is ${\bf v}_a (x,y)=f(x_a)f(y_a)\bar{{\bf v}}_a$ where  $f(x)=1-exp(-9x'^2)$ and $x' = 1-(2x-1)^2$.

\section{Fixed box: energy, vorticity and spin up}
\label{Sec:self-organization}
\subsection{General results}

In order to confirm that our SPH code behaves correctly it is convenient to compare our results  against the experimental and computational results of \citep{Clercxetal1998,Clercxetal1999,Clercxetal2001}\citep{Heijstetal2006,Maassenetal2002}. They showed that the energy and enstrophy decayed approximately as a power of $t$ until the flow is dominated by viscosity, and the angular momentum of the fluid may first increase in magnitude then slowly decay. This change in angular momentum is due to the reactive stress exerted on the fluid through its interaction with the walls.  To show that these phenomena are  also produced by our code we simulated turbulence  in a fixed square of side 1m. The turbulence was initiated  using $10\times10$ Gaussian vortices  and  $275\times275$ SPH particles and $\Re=2000$ as described in the previous section.  The SPH particles were  damped for $6000$ time steps after which the turbulence was initiated.

The vorticity $\omega_a$ for  particle $a$,  was determined by assuming that velocity  of the particles relative to particle $a$ was a linear function of the relative coordinates $x$ and $y$. The coefficients of this function were determined by least squares after which the derivatives of the velocity components were calculated. In the figures the vorticity field has been scaled between 0 and 1. The total enstrophy of the flow was then calculated according to
\begin{equation}
\xi( t)=\frac{1}{2}\sum_a\frac{m_a}{\rho_a}\omega_a^2.
 \label{enstrophy}
\end{equation}

\begin{figure*} 
\begin{center}
\begin{tabular}{p{4.7cm}p{4.7cm}p{5.1cm}}
\includegraphics[width=4.7cm]{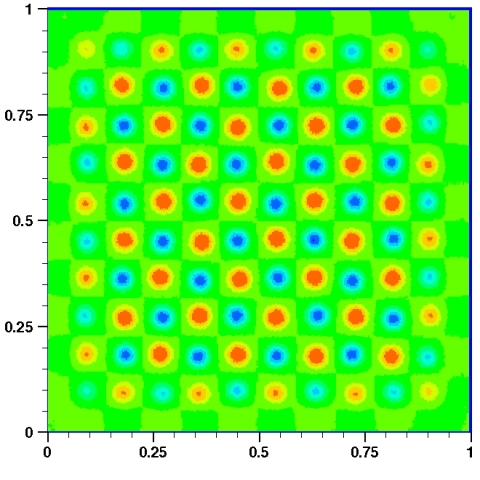}
&
\includegraphics[width=4.5cm]{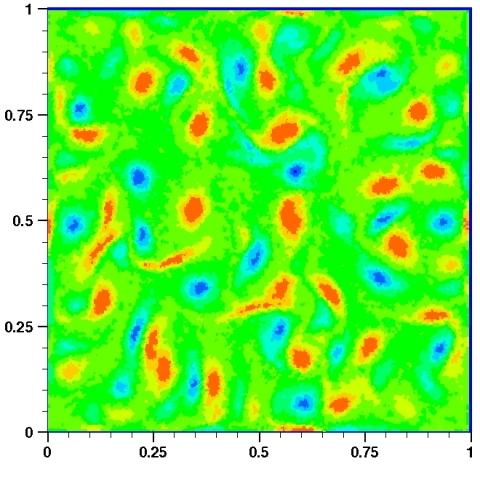}
&
\includegraphics[width=4.5cm]{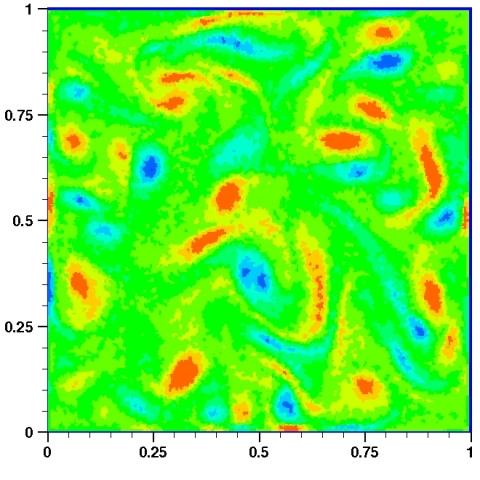}
\\
\includegraphics[width=4.5cm]{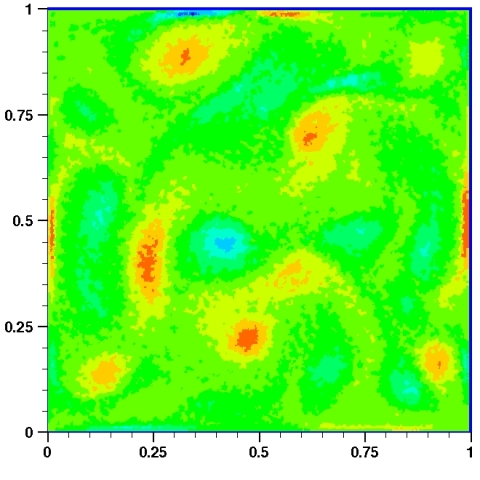}
&
\includegraphics[width=4.5cm]{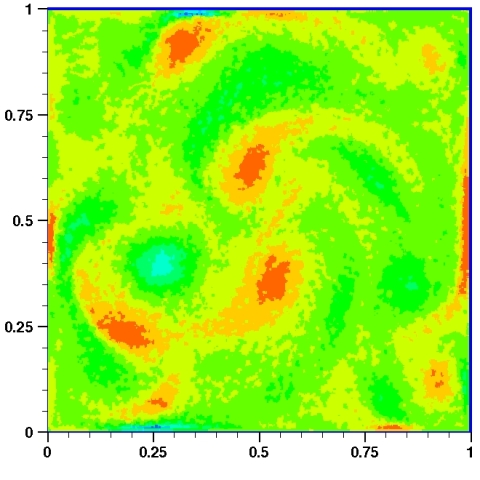}
&
\includegraphics[width=4.5cm]{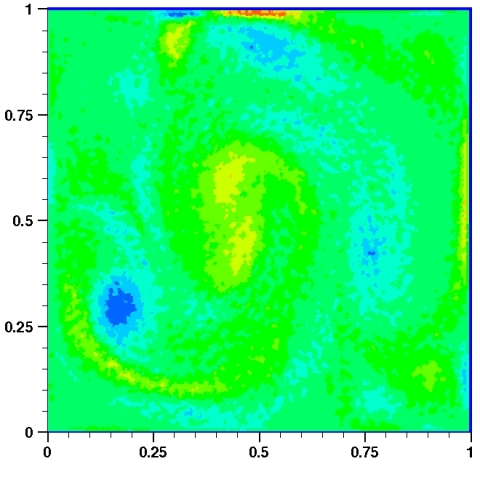}
\\
\includegraphics[width=4.5cm]{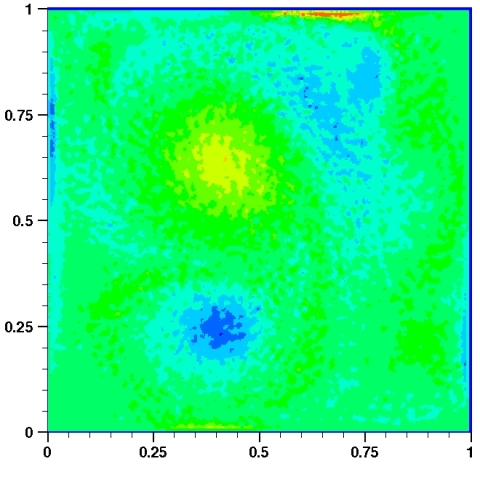}
&
\includegraphics[width=4.5cm]{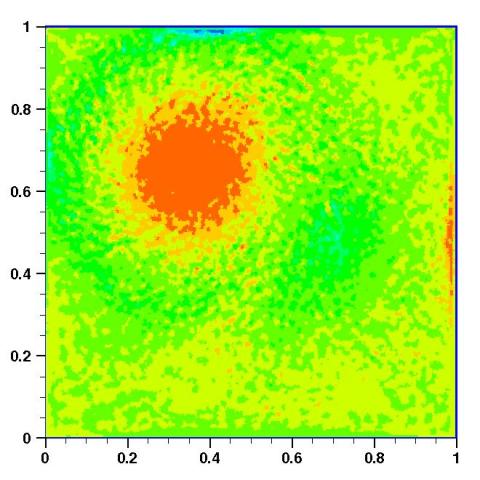}
&
\includegraphics[width=5.05cm]{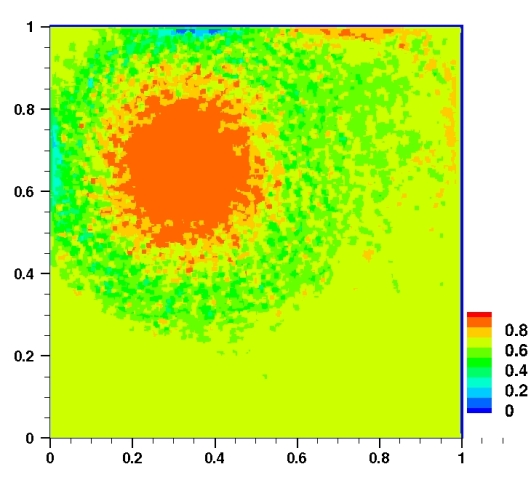}
\\
\end{tabular}

\caption{Vorticity plots of 2D decaying turbulence flow in a square container with no-slip boundary conditions ($\Re=2000$). The vorticity has been scaled between $0$ and $1$ in order to show the field at the later times. $t=0$,$8.08$,$13.5$,$27$,$35.14$,$48.7$,$73.1$,$146.5$ and $309.5$ respectively from left to right and top to bottom.}
\label{fig:spontaneous}
\end{center}
\end{figure*}

 Fig.~\ref{fig:spontaneous} shows that the average size of the vortices increases with time as  found  by \citep{Clercxetal1999} and \citep{Maassen2002}. During this time intense shearing occurs near the boundaries as can be seen, for example, in the first frame of the second row on the left and right sides. The  increase in the absolute value  of the fluid angular momentum (the spin-up) shown in the Fig.~\ref{fig:fixedrandom-angm}, accompanies  the formation of relatively large vortices. After spin-up, the angular momentum decays slowly, and this stage is characterised by a relaxation process to a monopolar structure that is more or less situated in the centre of the container. Subsequently, the flow relaxes viscously (the last frame in Fig.~\ref{fig:spontaneous}). 
 
 Figure 2(a) shows the variation of  kinetic energy for the present calculations (using $275 \times 275$ particles), those of  \citep{RobinsonMonaghan2011}, and those of \citep{Clercxetal1999} against scaled time  $t$ (time is scaled by $S/2U_{RMS}$,  where $U_{RMS}$ is the initial RMS velocity of particles and $S$ is the width of box).  The scaled time for the results of \citep{Clercxetal1999} are based on the best estimate of the time scaling used in their graphs.   All of these calculations indicate that $E_K \sim 1/t^{0.8}$ for $0 \lesssim t \lesssim 20$, with a more rapid decay for greater $t$. The agreement between the three sets of results is very satisfactory.  In figure 2(b) we show the results of the experiments and the present SPH results for the variation of the  normalized kinetic energy with time. The agreement is again very satisfactory. The decay of enstrophy is shown in figure 2(c).  The results for the different numerical codes  and the experimental results are similar to each other, and show an enstrophy decay that  is faster than the decay of the kinetic energy and given by $\sim 1/t^{1.4}$.  We can reasonably conclude that the decay of energy and enstrophy is not sensitive to the details of the initial velocity field (Gaussian vortices or a Chebyshev expansion), or sensitive to the two different treatment of boundaries in the SPH simulations.
 
\begin{figure*} 
\begin{center}
\begin{tabular}{p{5cm}p{5cm}p{5cm}}
\textbf {a}\includegraphics[width=4.8cm]{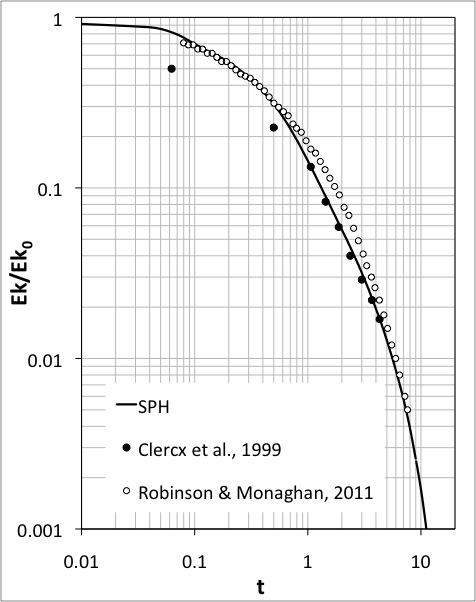}
&
\textbf {b}\includegraphics[width=4.8cm]{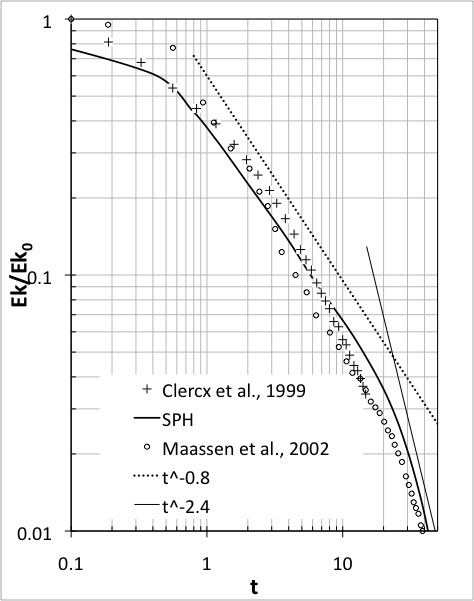}
&
\textbf {c}\includegraphics[width=4.8cm]{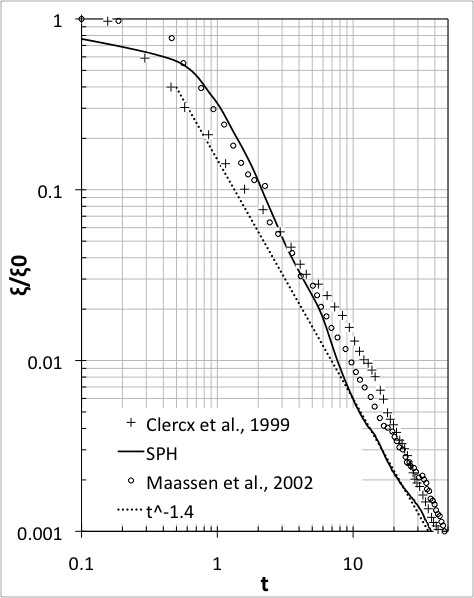}
\\
\end{tabular}

\caption{ ($\mathbf {a}$) The  normalized energy, $E/E\left(t=0\right)$ against scaled time for $\Re=1500$. Note that the results of the present calculation and those of Robinson and Monaghan (2011) and Clercx et al. (1999) are very similar. ($\mathbf {b}$) shows the decay of normalized kinetic energy for the present calculations and the experiments of Maassen et. al. (2002) for $\Re =2000$. Note the change in the maximum and minimum time.  The decay of the normalized enstrophy, $\xi /\xi(t=0)$ for $\Re=2000$, is  shown in (${\bf c}$) for the experiments, the present  simulations, and  those of Clercx et al. }
\label{fig:spont-angm}
\end{center}
\end{figure*}

\subsection{Convergence} 
\label{Sec:convergence}

\begin{figure*} 
\begin{center}
\begin{tabular}{p{6cm}p{6cm}}
$\bf a$\includegraphics[width=6.0cm]{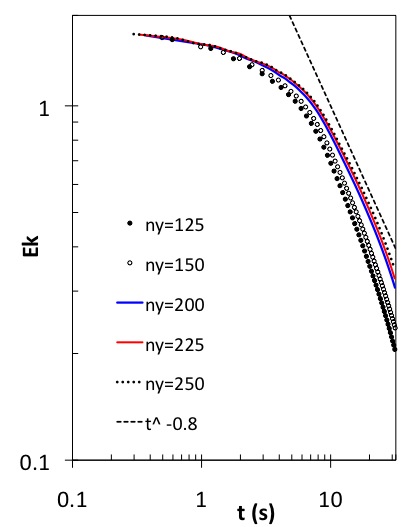}
&
$\bf b$\includegraphics[width =6.0cm]{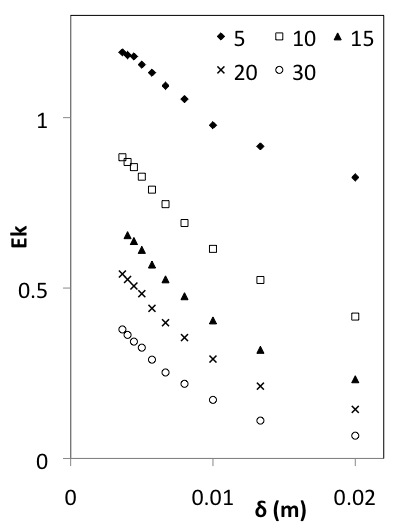}
\\
\end{tabular}
\caption{ The left frame shows the decay of kinetic energy  with time for different resolutions at $\Re=1000$. The initial particle spacing is given by $1/n_y$.  Once $n_y \ge 200$ the decay curves have a similar rate of decay. The right hand frame shows $E_k$ against the initial particle spacing for different times. The convergence is close to linear. }
\label{fig:convergence}
\end{center}
\end{figure*}

In Fig.~\ref{fig:convergence}{\bf a} shows the variation of $E_k$ with $t$ for different values of $n_y$ the number of particle spacings along a side of the boundary. The initial particle spacing is therefore  $1/n_y$.  The graphs for $E_k$ indicate that the results are converging.  In Fig.~\ref{fig:convergence}{\bf b} we show $ E_k$ at times 5, 10, 15, 20 and 25s as a function of resolution. These results indicate linear convergence. However, in order to determine the convergence of any quantity $A$ we follow \citep{Roache1997}, and assume
\begin{equation}
A(\delta) =A(0)+\beta\delta^p,
 \label{Euler}
\end{equation}
where $A(0)$ is the value at infinite resolution,  $\delta$ is the initial particle spacing, and $\beta$ and $p$ are parameters to be calculated.  The parameters $A(0)$, $\beta$ and $p$ can be found by evaluating $A$ at three values of $\delta$. The  trend of the kinetic energy with resolution, which is nearly linear, is shown in Fig.~\ref{fig:convergence}.

Fig.~\ref{fig:convergence}\--b, shows that for all times the change of kinetic energy with particle size, $\delta$, is close to linear for high resolutions. From the expression (\ref{Euler}) with $E_k$ for $n_y=100$, $200$ and $250$, we can calculate $E_k(0)$, $\beta$, and $p$ at other times , The resulting values are given in  Table 1. Using the calculated values of  $E_k(0)$, $\beta$, and $p$, it is possible to estimate $E_k$ at other times and other resolutions. Thus in Table I, the two columns under 275x275 show results calculated using SPH (the left column), and results using the convergence formula, denoted by $E_{est}$,  are shown in the right column. The agreement between the two columns is very good. In general the difference between the SPH values, and those estimated from the convergence formula, decreases  with increased resolution.
\begin{widetext}
\begin{table*}
\caption{Comparison of the convergence parameters, the SPH  values for kinetic energy together with the estimated values Est., from the  convergence study at $\Re=1000$. The resolution is given by the numbers at the head  of the 5th and later columns.}
\label{tab:convergence}
\begin{ruledtabular}
\begin{center}
\begin{tabular}{rccccccccccc}
t&$E_k\left(t\right)$&$-\beta$&$p$&\multicolumn{2}{c}{$275\times275$}&\multicolumn{2}{c}{$225\times225$}&\multicolumn{2}{c}{$150\times150$}&\multicolumn{2}{c}{$75\times75$}\\
\cline{5-6}\cline{7-8}\cline{9-10}\cline{11-12}&&&&SPH&Est.&SPH&Est.&SPH&Est.&SPH&Est.\\
\hline
5&1.255&247.4&1.475&1.192&1.193&1.179&1.171&1.093&1.103&0.957&0.831\\
10&1.05&36.33&0.962&0.884&0.886&0.855&0.851&0.746&0.756&0.524&0.476\\
15&0.845&29.96&0.917&0.666&0.671&0.638&0.636&0.527&0.542&0.319&0.272\\
20&0.730&20.04&0.831&0.541&0.541&0.506&0.506&0.398&0.417&0.212&0.173\\
30&0.632&6.786&0.585&0.379&0.377&0.343&0.346&0.252&0.269&0.111&0.088\\
\end{tabular}
\end{center}
\end{ruledtabular}
\end{table*}  
\end{widetext}

 Data presented in this table show that the kinetic energy converges less rapidly as the time increases.  and for longer times, the order of the convergence is less than one. The reason for the absence of second order convergence is not clear to us.  
 
 These convergence results indicate consistency 
\subsection{Angular momentum}

\begin{figure*} 
\begin{center}
\includegraphics[width=0.145\textwidth]{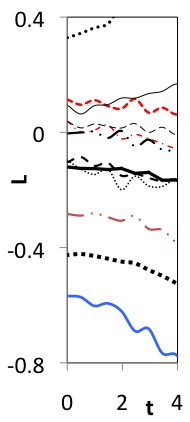}\includegraphics[width=0.6\textwidth]{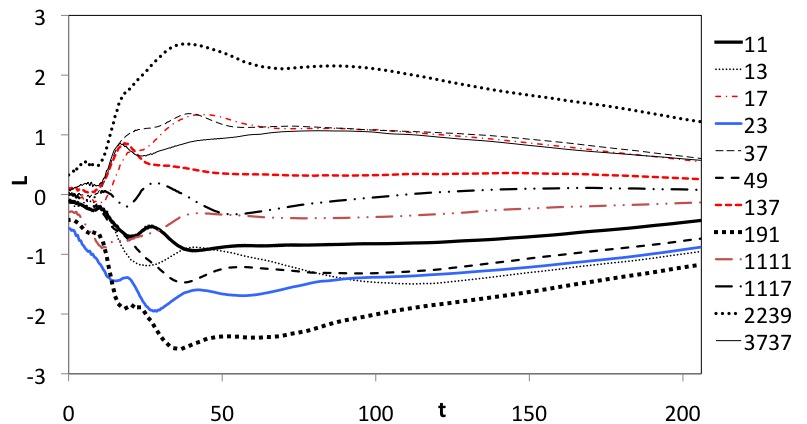}
\caption{The right frame shows the angular momentum against time (in second) for different random positions of the vortices forming the initial state in a fixed box at $\Re=1000$. The left frame shows a  magnified view of initial angular momentum. }
\label{fig:fixedrandom-angm}
\end{center}
\end{figure*}

\begin{figure*} 
\begin{center}
\includegraphics[width=0.4\textwidth]{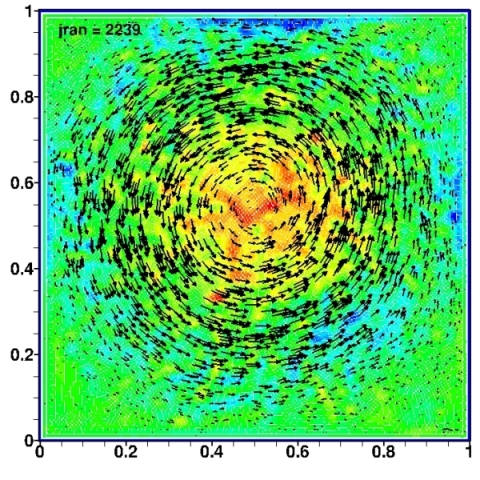} \includegraphics[width= 0.4\textwidth]{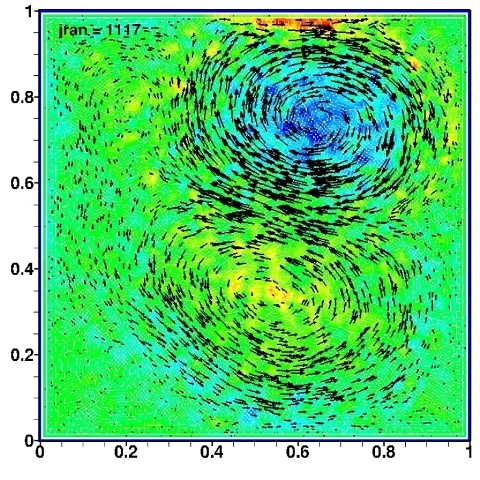}
\caption{Velocity field at $t=206 s$, for two different random initial setups in a fixed box at $\Re=1000$. Left frame $jran=2239$ shows a strong negative angular momentum, while the right frame calculated with $jran=1117$  shows negligible change in the angular momentum. }
\label{fig:fixedrandom-velocity}
\end{center}
\end{figure*}

As remarked earlier, a feature of the turbulence in two dimensions is the increase in the magnitude of the angular momentum, the spin-up. To confirm that our SPH simulation recovers the spin-up,  we used $4\times4$ vortices with $150\times150$ particles and $\Re=1000$. As noted earlier in $\S \ref{Sec:velocity field}$ we choose 12 different values of $jran$.  The time variation of the angular momentum is shown in Fig.~\ref{fig:fixedrandom-angm}. Among these $12$ simulations, $5$ show a rapid, spontaneous spin-up, and a further $4$ show spontaneous spin-down, which always characterised by the presence of a strong monopolar or a rotating tripolar structure. The other $3$ show no spin-up or very slow spin-up. During the intermediate stage of these runs a dipolar or quadrupolar structure is usually found. For example, the velocity field after 206 seconds is shown in Fig.~\ref{fig:fixedrandom-velocity}, for two different initial setups. This left frame of this figure shows the formation of a monopolar vortex which corresponds to the sudden increase in angular momentum. The right frame shows the formation of dipolar vortices, which corresponds to nearly zero angular momentum during the decay of the turbulent flow. \  \cite{Clercxetal1999} reported similar behaviour for an initial velocity field of Gaussian vortices or Chebyshev polynomials as did  \cite{RobinsonMonaghan2011} for Chebyshev polynomials.

Although the change in angular momentum is considerable when $jran$ is changed the corresponding difference  in the decay of kinetic energy with time is small, and all runs have a similar power law decay.



\subsection{Summary of the results with a fixed box}

Our SPH simulation recovers the principal phenomena found by other authors for turbulence in a two dimensional box with no-slip boundaries. This includes the growth of vortices, the decay of the kinetic energy and enstrophy, and the initial rapid increase in magnitude of the angular momentum followed by a slow decay. We conclude from this that our SPH code gives satisfactory results for two dimensional turbulence in a no-slip box and, in particular, the effectiveness of our boundary force model is confirmed.


\section{Freely rotating boxes}
\label{Sec:rotating}
\begin{figure*}
\begin{center}
\begin{tabular}{p{8.0cm}p{8.0cm}}
$\textbf{a}$ \includegraphics[width=7.0cm]{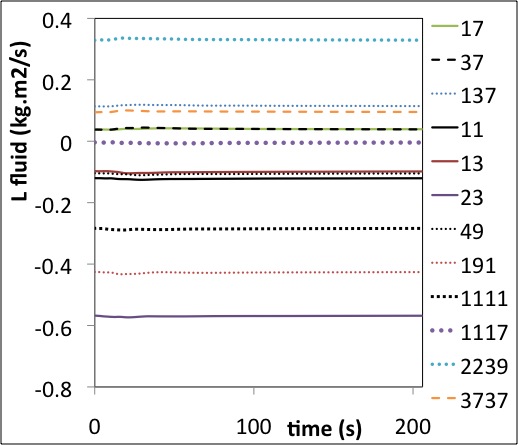}
&
$\textbf{b}$ \includegraphics[width=7.0cm]{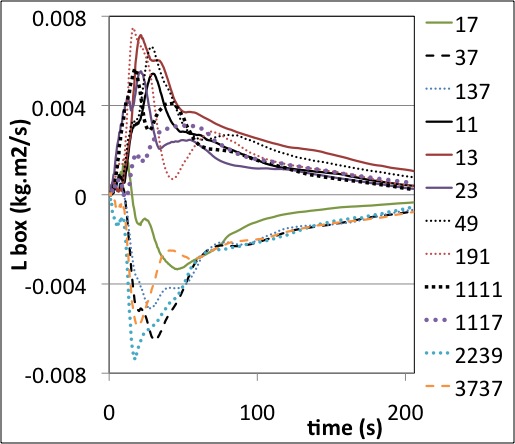}
\\
\end{tabular}
\caption{The effect of initial total angular momentum of fluid particles on  a freely rotating box at $\Re=1000$. 
$\textbf{a}$) total angular momentum of fluid particles, $\textbf{b}$)  total angular momentum of box particles. Total angular momentum of box plus total angular momentum of fluid particles is constant for each set. (the numbers in legend show $jran$'s in producing random numbers algorithm)}
\label{fig:freerandom-Lbox-lfluid}
\end{center}
\end{figure*}

\begin{figure*} 
\begin{center}
\begin{tabular}{p{8.0cm}p{8.0cm}}
\includegraphics[width=0.4\textwidth]{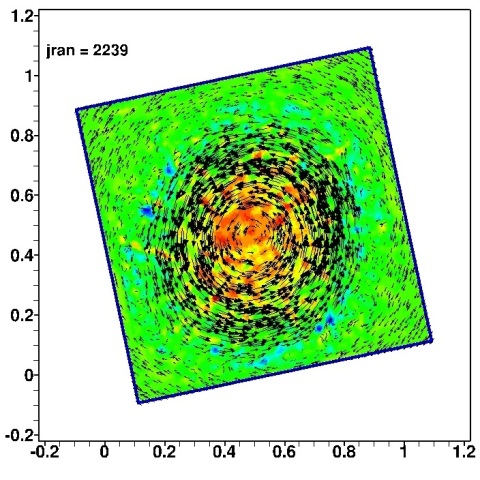} 
&
\includegraphics[width= 0.4\textwidth]{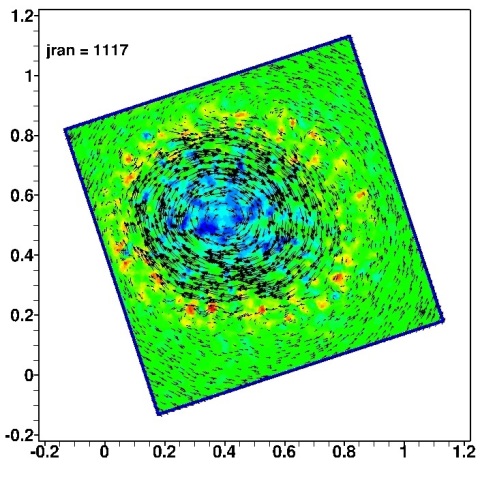}
\\
\end{tabular}
\caption{Velocity field at $t=206 s$, for two different random initial setups in a freely rotating box at $\Re=1000$. Left frame $jran=2239$ shows a strong negative angular momentum, while the right frame calculated with $jran=1117$  shows a strong positive angular momentum. }
\label{fig:freerandom-velocity}
\end{center}
\end{figure*}

We now consider the fluid angular momentum when the turbulence is initiated and the box is allowed to rotate under the torque produced by the fluid. In this case the angular momentum of the box and fluid is constant. In order to amplify the rotation of the box we give it a mass $ m_{box}=0.001\times {\rm (mass \ of \ fluid) }$ and moment of inertia $I_{\rm box} = \frac23 m_{box} S^2$.  The force ${\bf f}_j$ on boundary particle $j$ due to fluid particles is given by
\begin{equation}
{\bf f}_j = \sum_a m_j {\bf f}_{aj},
\end{equation}
and the torque on the box about its centre is 
\begin{equation}
 {\bf T} = \sum_j {\bf r}_{jo}\times  {\bf f}_j,
 \label{torquebox}
\end{equation}
where ${\bf r}_{jo}$ is the coordinate vector of boundary particle $b$, relative to the centre of the box. The angular velocity of box $\Omega$ is calculated from
\begin{equation}
I_{\rm box} \frac{d \Omega}{dt} = \bf T.
\end{equation}
The initial set up is same as the previous section.  Fig.~\ref{fig:freerandom-Lbox-lfluid} shows the angular momentum for the different values of $jran$.  The left frame of Fig.~\ref{fig:freerandom-Lbox-lfluid} shows the angular momentum of the fluid. The key point is that the fluid angular momentum remains nearly constant for all values of $jran$. The right hand frame of Fig.~\ref{fig:freerandom-Lbox-lfluid} shows the angular momentum of the box which may be positive or negative but the magnitude is always small and decreases with time.  The magnitude of the change in the fluid angular momentum can be calculated from the box angular momentum. In the case of the fixed box the fluid angular momentum typically changes by $\sim 1$, whereas in the present case it typically varies by 0.002, a factor 500 less.
The average size of the vortices increases with time as in the fixed box. For all 12 initial setups, a monopolar vortex formed at the centre of box. In contrast, some of the 12 setups  for the fixed box, produce  dipole or quadropole vortices, e.g. compare Fig.~\ref{fig:freerandom-velocity} with Fig.~\ref{fig:fixedrandom-velocity}. The decay of enstrophy in the freely rotating box is similar to that in the fixed box.

\begin{figure} 
\begin{center}
\begin{tabular}{p{11.5cm}}
\includegraphics[width=11.5cm]{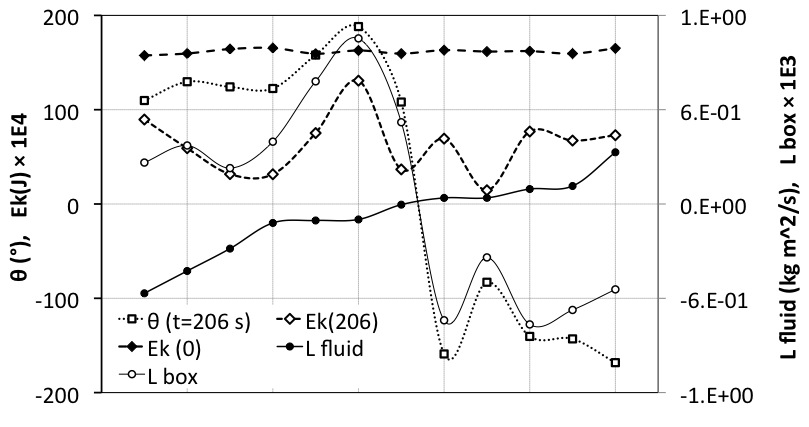}
\\
\end{tabular}
\caption{Initial total kinetic energy, $E_k(0)$, final total kinetic energy, total fluid angular momentum, $L_ {fluid}$, total box angular momentum, $L_{box}$, and total box rotation, $\theta$, after $206 s$ for a free\--rotating box at $\Re=1000$. The horizontal axis shows the number of initial setups for different random numbers, which are sorted according to $L_{fluid}$. }
\label{fig:freerandom-rotation}
\end{center}
\end{figure}

The decay of total kinetic energy of the fluid in the rotating box and the fixed box are very close, although, in the last stages of rotation, when the kinetic energy is  $\sim 10^{-3}$, the kinetic energy of the fluid in the rotating box is larger than the kinetic energy for the fixed box.
In Fig.~\ref{fig:freerandom-rotation} the results of different runs are sorted according to the initial angular momentum.  For each run the initial and final kinetic energy, final box angular momentum, and final box rotation are shown. The kinetic energy is multiplied by $10^4$ in this figure and the box angular momentum is multiplied by $10^3$. The direction of rotation of the box is given by the sign of initial angular momentum of fluid, but the magnitude of box angular velocity is not related to the magnitude of the initial angular momentum. The initial kinetic energy for all runs is nearly same, but the  kinetic energy after $206$ seconds can be significantly different according to the initial set of random shifts of the vortices. 

\section{The effect of Background rotation}

In the previous cases the initial angular velocity $\Omega_0$ of box and fluid before inserting the Gaussian vortices was zero. Another interesting case that is relevant to turbulence in the earth's atmosphere or ocean, is the behaviour of turbulence when the fluid has a background rotation.  Forced quasi-two-dimensional turbulence in rotating containers has been studied both by experiment and by numerical simulations \citep{Clercxetal2005,Molenaaretal2004}. By adding an oscillation to the rotation of the container it is possible to determine how  shear near the boundary generates vorticity. The vorticity generated in this way subsequently moves into the central regions of the container though in the SPH simulations of \citep{RobinsonMonaghan2011}, the vorticity generated in this way dissipated more rapidly than in the calculations of \citep{Molenaaretal2004}.

In this section we study the turbulence when  the fluid and box are set  rigidly rotating with angular velocity $\Omega_0$, after which the vortices are added. The vorticity of the initial state is then $\omega_0 = 2\Omega_0$. The additional vorticity due to the gaussian vortices satisfies the usual vorticity diffusion equation for two dimensions (equation~\ref{vort}).
If the turbulence is studied in the  frame rotating with angular velocity $\Omega_0$ the centrifugal force is balanced by the radial pressure gradient.  The Coriolis force, $2 (\Omega_0 {\bf \hat z} ) \times {\bf v}$, changes the motion of the flow arising from the vortices.

The previous Gaussian vortices were used with $jran=11$, and the rotation was chosen so that the Rossby numbers, $Ro=2U_{rms}/ \Omega_0S$ were in the range $0.57 \le R_o \le 11.0$.   $U_{rms}$ is the root mean square velocity of vortices, which is equal to $0.057$, and $S=1$ is the width of box. The results for infinite $Ro$ were discussed in the previous section. Here $Ro\simeq11$, $1.1$, and $0.57$ equivalent to $\Omega_0=0.01$, $0.1$, and $0.2$ respectively. The Reynolds number was set to 1000 for all studies hereafter. 

After damping, the box and fluid were rotated with $\Omega_0$ for $2000$ time steps after which the box and fluid were rotating rigidly. The Gaussian vortices were then added to the velocity field.  The subsequent velocity and  vorticity field then depends on the rotation of the box and how the added vortices change in response to the background rotation. We consider two cases. The first, denoted by A, is where the rigid rotation of the box is maintained. The second, denoted by B, is when the box is allowed to respond freely to the fluid stresses on it.  In the following we denote the kinetic energy and the enstrophy in the inertial frame by $E_k$ and $\xi$ respectively, and in the rotating frame  by $\hat E_k$ and $\xi_{rot}$.

\subsection{The box with maintained rigid rotation}
\begin{figure*} 
\begin{center}
\begin{tabular}{p{8.0cm}p{8.0cm}}
\includegraphics[width=8.0cm]{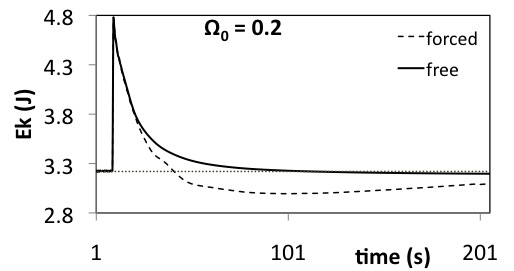}
&
\includegraphics[width=8.0cm]{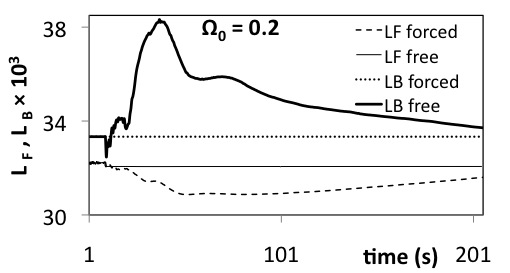}
\end{tabular}
\caption{Total kinetic energy, $E_K$ (left) and total angular momentum of box, $L_B$ and fluid, $L_F$ (right) for different background angular velocities, $\Omega_0$. "forced"  denotes the case of box with maintained rigid rotation, and "free" indicates that the box is allowed to rotate freely after the vortices are initiated. In the right hand frame the dashed and thin continuous lines denote the angular momentum of the fluid for the forced and freely rotated box respectively, and the dotted and thick solid lines denote the angular momentum of the box for the forced and freely rotated box respectively, calculated in the inertial frame.}
\label{fig:Ek-L rotating}
\end{center}
\end{figure*}

 In this case the box rotates rigidly with angular velocity $\Omega_0$. The total kinetic energy of box and fluid is shown in the left frame of Fig. ~\ref{fig:Ek-L rotating} for $\Omega_0=0.2$. The dashed line shows the kinetic energy for the rigidly rotating box. The kinetic energy increases suddenly because of the extra kinetic energy of the vorticity field. This extra kinetic energy decays rapidly to a steady state where the fluid and box are rotating rigidly.  The  relative decay of the kinetic energy, $E_k$, due to the vortices at different $\Omega_0$'s is calculated by $\Delta E_k =(E_k-E_{in})/\hat{E_{in}}\times100$, where $E_{in}$ is the initial kinetic energy before adding vortices, and $\hat{E_{in}}$ is the total initial kinetic energy of vortices. The kinetic energy of the  fluid in the inertial frame is
 \begin{equation}
 E_k=\hat{E_k} + \frac{1}{2}I\Omega_0^2 + \Omega_0\hat{L},
 \label{energy}
\end{equation}
where, $\hat{E_k}$ and $\hat{L}$ are the kinetic energy and angular momentum of the fluid in the rotating frame, respectively, and $I$ is the moment of inertia of the fluid around the box centre. $\hat{E_k}$ and $\hat{L}$ are zero before adding vortices. 

The calculated kinetic energy in the rotating frame is the same for all $\Omega$'s, for a given set of vortices. The values of $\Delta E_k$, using $E_k$ as the minimum kinetic energy which occurs nearly at time 100s,  is shown in column $7$ of table~\ref{tab:decay}. The results in table~\ref{tab:decay} show that the rate of decay of the kinetic energy of the fluid increases significantly by increasing the box angular velocity.  Therefore, while the rate of change of kinetic energy in the rotating frame is always negative, the rate of change of kinetic energy in the inertial frame can be negative or positive, due to the last term in equation~\ref{energy}. This can be seen easily from the left frame of Fig.~\ref{fig:Ek-L rotating}. 
\begin{table}
\caption{The rate of decay in kinetic energy, $\Delta E_k$, total angular momentum, of box $L_{Box}$, and fluid, $L_{Fluid}$ for when the box continues to rotate rigidly at different angular velocities, $\Omega_0$. $\hat E_{in}$ is the initial kinetic energy of vortices.}
\label{tab:decay}
\begin{ruledtabular}
\begin{tabular}{rcccccc}
&&\multicolumn{3}{c}{$L_{Fluid}$}&&\\
\cline{3-5}
(1)&(2)&(3)&(4)&(5)&(6)&(7)\\
$\Omega_0$&$L_{Box}$&$Min$&$Max$&$\Delta L$&$\hat {E_{in}}$&$\Delta E_k (t=100)$\\
\hline
0.00&0.000&-0.94&0.00&0.94&1.575&+0.092\\
0.01&0.002&0.414&1.61&1.20&1.575&-0.202\\
0.10&0.017&14.84&16.10&1.26&1.573&-6.331\\
0.20&0.033&30.87&32.21&1.35&1.575&-14.63\\
\end{tabular}
\end{ruledtabular}
\end{table}  

\begin{figure*} 
\begin{center}
\begin{tabular}{p{6.5cm}p{6.5cm}}
\includegraphics[width=6.2cm]{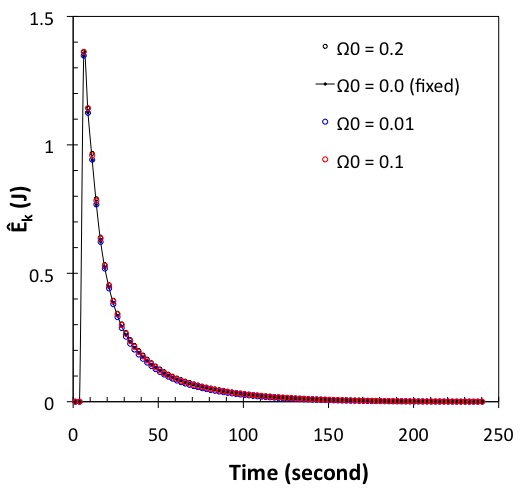} 
&
\includegraphics[width=6.0cm]{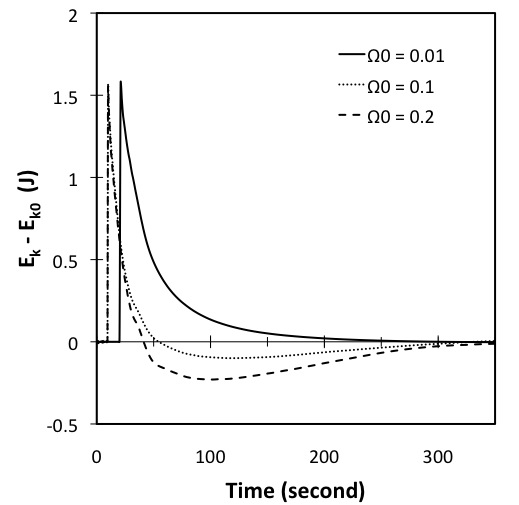} 
\end{tabular}
\caption{The left frame shows the decay of the kinetic energy in the rotating frame for a rigidly rotating box. As conjectured it is independent of the angular velocity. The right frame shows the change in the kinetic energy with time as measured in the inertial frame.}
\label{fig:kinenergyangvel}
\end{center}
\end{figure*}

The right frame of Fig.~\ref{fig:Ek-L rotating} shows the total angular momentum of the box and the fluid for the case where the box rotates rigidly. The angular momentum of fluid in the inertial frame decreases to a minimum (column 3 of Tab.~\ref{tab:decay}), then it decays gradually until at infinity it reaches the initial angular momentum of fluid before adding Gaussian vortices (column 4 of Tab.~\ref{tab:decay}). Recall that in this section the box rotates rigidly throughout the simulation. The change in the angular momentum of the fluid is almost independent of  $\Omega_0$, e.g the maximum change in the angular momentum of fluid is shown in column 5 of Tab.~\ref{tab:decay}, which shows the subtraction of column 3 from column 4. In the left frame of Fig.~\ref{fig:kinenergyangvel} we show the decay of kinetic energy in the rotating frame for different values of the angular velocity of the box. As conjectured  after (9) it is expected that the decay of the kinetic energy calculated in the rotating frame should be independent of the rotation because the equations for the decay are identical to those in a fixed box.The right hand frame of Fig.~\ref{fig:kinenergyangvel} shows that the decay of $E_K$ varies significantly with the angular velocity of the box.

\begin{figure*} 
\begin{center}
\begin{tabular}{p{8.1cm}p{8.1cm}}
$\textbf{a}$\includegraphics[width=7.8cm]{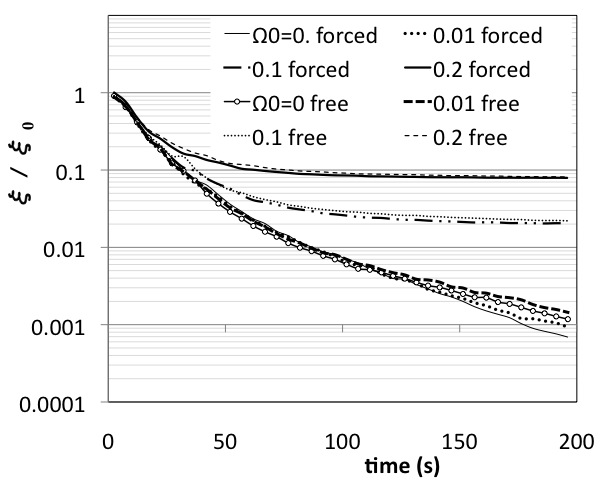}
&
$\textbf{b}$\includegraphics[width=7.8cm]{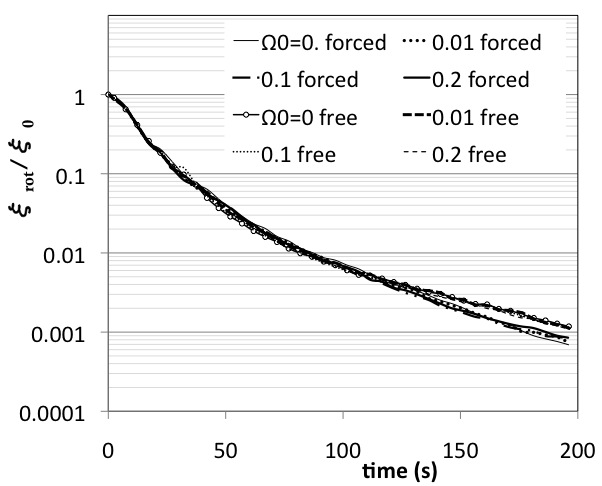}
\\
\end{tabular}
\caption{The effect of angular velocity, $\Omega_0$, on enstrophy, $\xi$, $\bf a)$ total enstrophy, calculated in inertial frame, $\bf b)$ net enstrophy due to Gaussian vortices, calculated in rotating  frame. "forced"  denotes the case of box with maintained rigid rotation, and "free" indicates that the box is allowed to rotate freely after the vortices are initiated.  $\xi_0$ is the enstrophy of initial Gaussian vortices equal to 0.999. }
\label{fig:enstrophy-vortices}
\end{center}
\end{figure*}

The enstrophy calculated in the rotating frame, $\xi_{rot}$, is independent of  $\Omega_0$ and is same as the fixed box, this is shown in the right frame of Fig.~\ref{fig:enstrophy-vortices}. Using the no-slip boundary conditions it can be shown that the enstrophy $\xi$ calculated in the inertial frame is  8$\Omega_0^2S^2$ larger than $\xi_{rot}$, where $S$ is the half-width of box. As a consequence, when the enstrophy in the rotating frame has decayed to zero, $\xi = 8\Omega_0^2S^2$ as shown in the left frame of Fig.~\ref{fig:enstrophy-vortices}(graphs labelled 'forced '). 

\subsection{The box freely-rotating after the vortices are initiated}

 In this case the box rotates freely after the vortices are initiated in the rigidly rotating fluid. The subsequent motion of the box is due to the  stresses on it from the fluid. This class of  problem is relevant to a wide range of engineering fluid-structure interaction problems such as vortex-induced vibrations. The total kinetic energy of box and fluid in the inertial frame, $E_k$, for this case is shown in the left frame of Fig.~\ref{fig:Ek-L rotating}.  By adding Gaussian vortices to the background velocity field, a sudden rise in kinetic energy occurs which is followed by a sharp decline to the background kinetic energy. The kinetic energy is calculated at the same times that the kinetic energy of the box with maintained rigid rotation is minimum for each $\Omega_0$. The relative decay is shown in the last column of table~\ref{tab:decay in rotation freely rotation}.   
The decay rate of total kinetic energy (liquid and box) when the  box is freely rotating is less than the when it is rigidly rotated, and this difference increases significantly by increasing the box angular velocity (see Tab.~\ref{tab:decay} and Tab.~\ref{tab:decay in rotation freely rotation}), around $15 \%$ for $\Omega_0=0.2$. The decay rate of the kinetic energy increases as the angular velocity increases for both the freely rotating box and the rigidly rotating box.\\
 
\begin{table}
\caption{The rate of decay of kinetic energy, $\Delta E_k$, total angular momentum, of box $L_{Box}$, and fluid, $L_{Fluid}$ for the case of a box  which is rotated initially with $\Omega_0$ then rotates freely after the  vortices are initiated.}
\label{tab:decay in rotation freely rotation}
\begin{center}
\begin{ruledtabular}
\begin{tabular}{cccccccc}
&\multicolumn{3}{c}{LBox}&\multicolumn{3}{c}{LFluid}&\\
\cline{2-4}\cline{5-7}
(1)&(2)&(3)&(4)&(5)&(6)&(7)&(8)\\
$\Omega_0$&$Min$&$Max$&$\Delta L$&$Min$&$Max$&$\Delta L$&$\Delta E_k (t=100)$\\
\hline
0.00&0.0&0.005&0.0054&-0.126&0.0&0.126&+1.84\\
0.01&0.002&0.007&0.0051&1.484&1.609&0.126&+1.67\\
0.10&0.017&0.022&0.0053&15.963&16.107&0.144&+1.46\\
0.20&0.033&0.038&0.0050&32.060&32.268&0.208&+0.22\\
\end{tabular}
\end{ruledtabular}
\end{center}
\end{table}  

The right frame of Fig.~\ref{fig:Ek-L rotating} shows the total angular momentum of box and the total angular momentum of fluid, when the box is free to rotate after the gaussian vortices are initiated. The angular momentum of the box is not constant and increases to a maximum, then it decays gradually until it reaches the angular momentum of the box before initiating the Gaussian vortices.  Note that the combined angular momentum of fluid and box is constant and the angular momentum of the box is scaled up by a factor $10^3$. The  increase in total angular momentum of the box is not related to the $\Omega_0$ (column 4 of Tab.~\ref{tab:decay in rotation freely rotation}). The initial angular momentum of vortices is around -0.126. The thin solid line in the right frame of Fig.~\ref{fig:Ek-L rotating}, which shows the total angular momentum of the fluid in the inertial frame, shows a small drop by -0.126 shortly after $t=1$, then remains nearly constant, varying slightly as the box angular momentum changes. These changes in the angular momentum are indicated in columns 5 to 7 of Tab.~\ref{tab:decay in rotation freely rotation}. Thence the angular momentum of the fluid remains approximately constant for each $\Omega_0$.\\

From the right frame of Fig.~\ref{fig:enstrophy-vortices}, and focussing on the graphs labelled 'free', it can be seen that the decay of enstrophy is independent of $\Omega_0$ as in the case of the rigidly rotating box. The enstrophy of the fluid when the box is freely rotated is always slightly larger than when it is rotating rigidly. The right frame of Fig.~\ref{fig:enstrophy-vortices} shows the decay of the enstrophy after subtracting the initial background rotation from the total flow. Although the rotation of the box can change when it is freely rotating it is convenient to consider the enstrophy in the frame rotating with the initial angular velocity. The enstrophy shown in the right frame of Fig.~\ref{fig:enstrophy-vortices} is calculated in this frame both for the forced and free rotations.  Our comparison shows that the  enstrophy decay in this frame is nearly same for all $\Omega_0$'s, and is very close (0.17 \%) to that of the rigidly rotated box .

\begin{figure*} 
\begin{center}
\begin{tabular}{p{8.0cm}p{8.0cm}}
$\textbf{a}$\includegraphics[width=7.8cm]{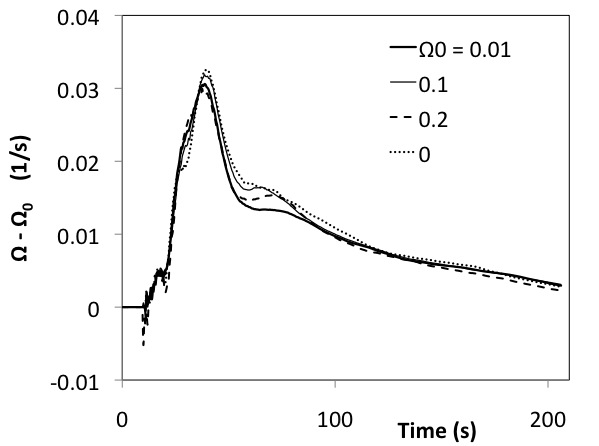}
&
$\textbf{b}$\includegraphics[width=7.8cm]{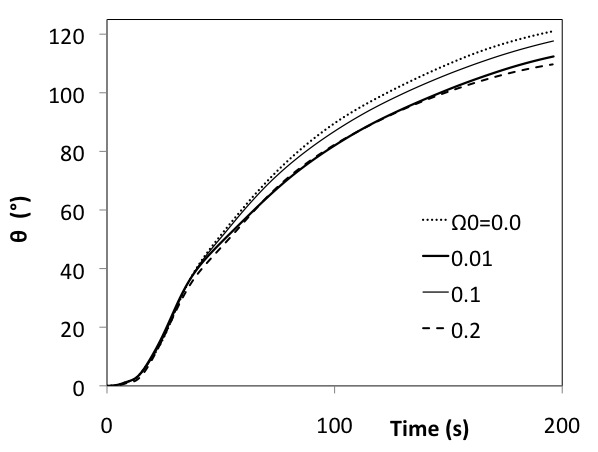}
\\
\end{tabular}
\caption{The effect of background rotation, $\Omega_0$, on the box rotation. The box is free to rotate after the gaussian vortices are initiated; $\textbf{a}$) The box angular velocity due to vortices, $\Omega - \Omega_0$, versus time, $\textbf{b}$) The net rotation angle due to the vortices versus time, which is calculated after subtracting the forced rotation angle, $\Omega_0. t$, from the free rotation angle.}
\label{fig:theta-vortices}
\end{center}
\end{figure*}

The angular velocity of the box due to the vortices is shown in the left frame of Fig.~\ref{fig:theta-vortices}. This figure shows that the effect of vortices is nearly the same for all background rotations and, as a consequence, the total rotation of the box is similar for all values  of  $\Omega$ as shown in Fig.~\ref{fig:theta-vortices} for different $\Omega_0$'s. 

\subsection{Summary of the results with Background rotation}

\begin{figure*} 
\begin{center}
\begin{tabular}{p{5.3cm}p{5.3cm}p{5.3cm}}

\includegraphics[width=5.0cm]{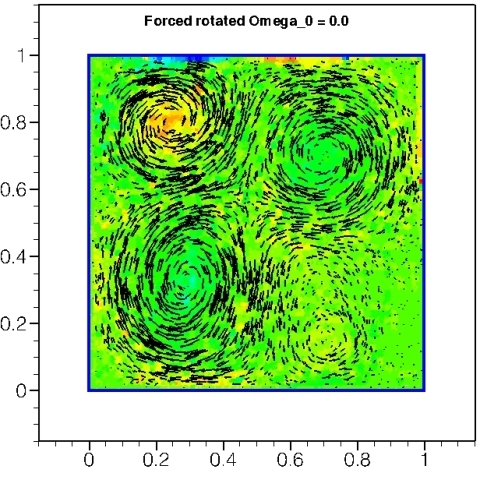}
&
\includegraphics[width=5.0cm]{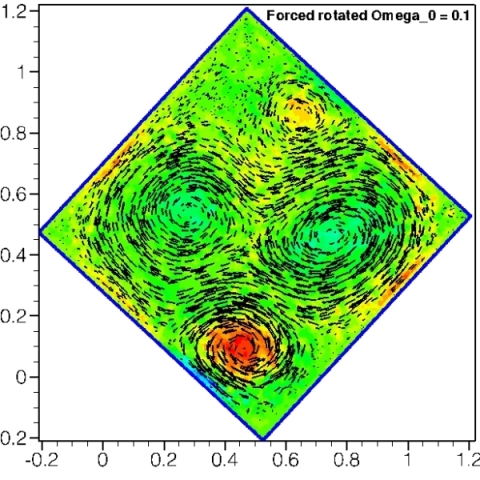}
&
\includegraphics[width=5.0cm]{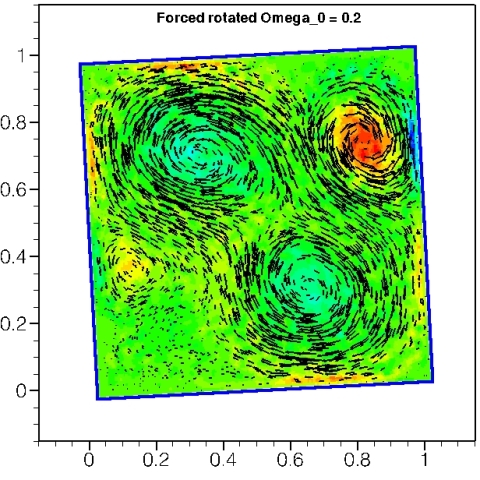}
\\
\includegraphics[width=5.0cm]{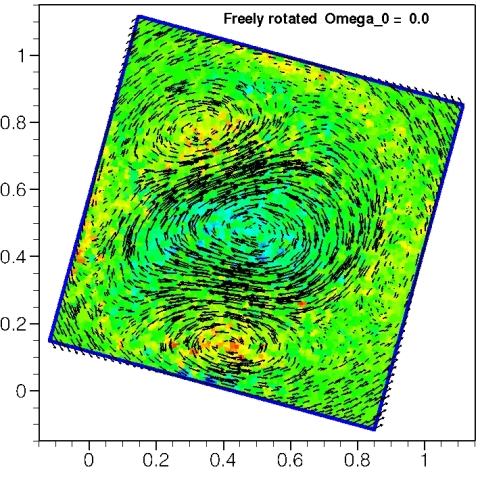}
&
\includegraphics[width=5.0cm]{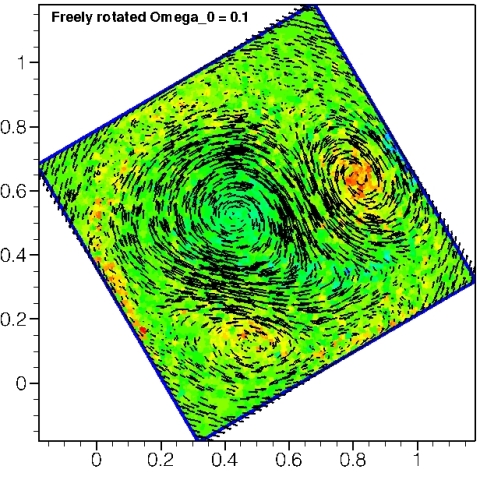}
&
\includegraphics[width=5.0cm]{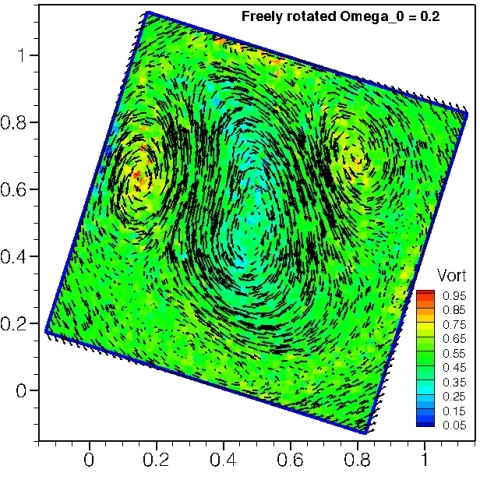}
\\
\end{tabular}
\caption{Vorticity and velocity field for different angular velocities, first, second and third columns from the left show results at time=76s after adding vortices for $\Omega_0= 0.0, 0.1, $ and 0.2, respectively. 
The top row show the box with maintained rigid rotation, and the bottom row show the freely rotating box results. All velocity fields are calculated the in rotating frame.}
\label{fig:Vort-free-rigid}
\end{center}
\end{figure*}

In the case of the box with maintained rigid rotation, if the flow is considered in a rotating frame, the effect of background rotation on the flow characteristics like vorticity structure (see Fig.~\ref{fig:Vort-free-rigid}), enstrophy, kinetic energy of vortices, and angular momentum of the fluid is negligible.  These results are consistent with the fact that the equations for the decay of enstrophy and kinetic energy in the rotating frame are identical to those for a fixed box. The decay rate of the kinetic energy in the inertial frame increases with $\Omega_0$. 

When the box is allowed to rotate freely, the effect of the background rotation is negligible, though the vortex structure is completely different from that for the box with maintained rigid rotation (see Fig.~\ref{fig:Vort-free-rigid}). 

\section{Conclusion}
\label{conclusion}

We have described the results of numerical simulations of decaying two-dimensional turbulence inside a rotating square container with rigid no-slip boundaries, using SPH. Three cases were studied here and, in all cases, the turbulence was initiated by superimposing the velocity field of a set of  gaussian vortices.  The first two cases involve a fixed box, and a box that is free to rotate under the stresses from the fluid. The third involves adding gaussian vortices to a box and fluid  rotating rigidly.  After which the box was either forced to maintain its rigid rotation, or allowed to rotate under the stresses from the fluid. 

Our simulations confirm that our SPH code correctly simulates turbulent flows in a fixed square box with no-slip boundaries. In particular, the simulations show a rapid self-organisation of the flow toward one or two large vortices as the total kinetic energy of the flow decreases.  By changing the random shift of the initial vortices from a regular grid the initial conditions can be changed. The resulting simulations provide an ensemble from which more general properties can be deduced. It was observed that 5 of them show a rapid spontaneous spin-up, and a further 4 showed spontaneous spin--down. All of them were accompanied by an increase in the magnitude of the angular momentum of the fluid. That behaviour was always accompanied by a strong monopolar or a rotating tripolar structure. The other 3 simulations showed no spin-up or very slow spin-up.  During the intermediate stage of these simulations  a dipolar or quadrupolar structure was usually found, and the net angular momentum of flow remains approximately constant or decayed only very slowly.  The decay of the total kinetic energy of the flow was more rapid in all cases. These results are in good agreement with other experimental and numerical studies \citep{Clercxetal1999, Maassenetal2002,RobinsonMonaghan2011}. 

When the box was allowed to rotate under the stresses from the fluid the angular momentum of the fluid remained approximately constant and the total angular momentum of box and fluid remained constant as it should. Simulations with slightly different initial conditions  change the dynamics of the box. In particular, the  box angular momentum changes and therefore the amount and direction of its rotation. All 12 initially different setups concluded in a  strong monopolar vortex. The rate of decay of the kinetic energy is nearly the same as that for the box with maintained rigid rotation. 

From the study of the effect of a background rotation on the decaying turbulence, it was seen that for the box with maintained rigid rotation, the effect of box angular velocity on the angular momentum, vorticity structure, velocity field, kinetic energy, and enstrophy is negligible, when these quantities are calculated in a frame rotating with $\Omega_0$. This is consistent with the fact, already noted, that the equations for the decay of kinetic energy and enstrophy in the rotating frame are the same as in the inertial frame. For the box freely\--rotating after the turbulence was initiated, kinetic energy, angular momentum,vorticity structure, velocity field, enstrophy, and rotation, calculated in the rotating frame were independent of $\Omega_0$. Although some parameters like kinetic energy or enstrophy calculated in the rotating frame are similar for both the freely rotated box and the box with maintained rigid rotation, the vorticity structure, velocity field and angular momentum are completely different for these two situations.

The resolution required for these SPH simulations depends on the accuracy required and the time for which that accuracy should be maintained.  The results of this paper and those of \citep{RobinsonMonaghan2011}, and \citep{Monaghan2011}, show that for the square box of half width $S$ the number of particles should be $\sim 200$ with particle spacing $dp = 2S/200$, in order to determine the energy decay to within $10\%$ for $t<15$. The resolution length can by estimated from $h = 1.5 dp$ and this is $ \sim 0.5S/\sqrt{\Re} $ which agrees with the estimate of \citep{Clercxetal1999}. We also note that our  treatment of the boundaries using boundary forces gives results in good agreement with those of \citep{RobinsonMonaghan2011} for the energy decay obtained using a boundary modelled by layers of fixed fluid particles.  However, the convergence of our results for the energy decay, obtained using the Wendland kernel, appear to be faster than those of \citep{RobinsonMonaghan2011} (see for example their figure 9) using the cubic spline kernel. This agrees with the experience of Robinson (private communication) who repeated some of his calculations using the Wendland kernel instead of the cubic spline kernel.

In addition to the class of problems considered here SPH can be applied to other turbulent flows where it has many advantages.  One of these is the turbulent flow produced by physical stirrers, for example cylindrical rods moving on specified paths. Such a problem can be simulated easily with SPH and, in unpublished work, we have studied the turbulence produced by such stirrers moving on a variety of trajectories.  SPH has also proven useful for studying breaking waves especially those formed by a flow hitting and running up a wall and finally forming a backward breaking wave. A region of strong turbulence is created by the impact of the breaking wave on the incoming fluid.  This problem, which is a key feature in sloshing in marine tanks, could be tackled using an SPH code of the kind we have described. 
 
\section{Acknowledgments}
This research was funded with the support of ARC Discovery grant DP0881447 (Analysis of two\--phase sloshing in marine tanks).
\section{References}
\bibliographystyle{unsrt}
\bibliography{references}

\end{document}